\documentclass[a4paper,fleqn,usenatbib]{mnras}

\usepackage{newtxtext,newtxmath}

\usepackage[T1]{fontenc}
\usepackage{ae,aecompl}

\usepackage{graphicx}	
\usepackage{amsmath}	
\usepackage{amssymb}	

\title[Resonance Locations in NGC 613]{Determination of resonance locations in NGC 613 from morphological arguments}

\author[M.~S.\ Seigar et al.]{
Marc S. Seigar,$^{1}$\thanks{E-mail: msseigar@d.umn.edu (MSS)}
Amber Harrington,$^{2}$
and Patrick Treuthardt$^{3}$
\\
$^{1}$Department of Physics \& Astronomy, University of Minnesota Duluth, 1023 University Dr, Duluth, MN 55812, USA\\
$^{2}$Department of Physical Sciences, Arkansas Tech University, 1701 N.\ Boulder Ave, Russellville, AR 72801, USA\\
$^{3}$Astronomy and Astrophysics Research Lab, North Carolina Museum of Natural Sciences, 11 W. Jones Street, Raleigh, NC 27601, USA
}

\date{Accepted XXX. Received YYY; in original form ZZZ}

\pubyear{2018}

\begin{document}
\label{firstpage}
\pagerange{\pageref{firstpage}--\pageref{lastpage}}
\maketitle

\begin{abstract}
  In this paper, we present $BVRI$ imaging data of NGC 613.  We
  use these data to determine the corotation radius of the bar, using the
  photometric phase crossing method.  This method uses the phase angle of the
  spiral structure in several wavebands, and looks for a crossing between the
  blue ($B$) light and the redder wavebands (e.g., $R$ or $I$).  For
  NGC 613, we find two phase crossings, an outer phase crossing
  at $136\pm8$ arcsec and an inner phase crossing at $16\pm8$ arcsec.  We
  argue that the outer phase crossing is due to the bar corotation radius,
  and from the bar length of $R_{\rm bar}=90.0\pm4.0$ arcsec
  we go on to calculate
  a relative bar pattern speed of $\mathcal{R}=1.5\pm0.1$, which is consistent
  with the results of previous methods described in the literature.  For a
  better understanding of the inner phase crossing, we have created
  structure maps in all four wavebands and a $B-R$ color map.  All of our
  structure maps and our color map highlight a nuclear ring of star formation
  at a radius of $\sim4$ arcsec, which had also been observed recently
  using ALMA.  Furthermore, the radius of our inner phase crossing
  appears to be consistent with the size of a nuclear disk of star formation
  that has been recently detected and described in the literature.  We
  therefore suggest that the phase crossing method can be used to detect
  the size of nuclear star formation regions as well as the location of
  corotation resonances in spiral galaxies.
\end{abstract}




\section{Introduction}
\label{Intro}

Identifying corotation resonances and their locations in barred spiral
galaxies is important for a number of reasons.  First, resonances are a
requirement of quasi-stationary density waves, so finding them would be
evidence that supports density wave models.  Furthermore, the location of
the bar corotation resonance is related to how it transfers material from
the outer parts of the galaxy to the central regions, i.e., the process
that dominates secular evolution in nearby disk galaxies.  Also, resonances
scatter stars and cause disk heating.
In the determination of bar pattern speeds, authors often determine
the dimensionless quantity $\mathcal{R}$, which is
the ratio of the bar corotation radius $R_{\rm CR}$ divided by the bar radius
$R_{\rm bar}$, i.e., $\mathcal{R}=R_{\rm CR}/R_{\rm bar}$.

Several other methods that can be used to measure the location of the corotation
resonance (CR) and/or bar pattern speeds.
This includes the Tremaine-Weinberg method \citep[][hereafter TW]{TW84}, which
is the most direct method of determining the CR location
\citep[see, e.g.][]{A15,TBSL07}.  The TW method requires a significant
amount of telescope time and it is only useful for
galaxies where the bar major axis and galaxy major axis are offset by
$20^{\circ}-70^{\circ}$ \citep{GKM03} and for galaxies that are inclined
by $50^{\circ}-60^{\circ}$ \citep{D03}, so there are a
very limited number of galaxies to which TW can be applied.

Other, alternative methods to TW include computer simulations or response
models \citep[e.g.][]{RSL08,TSRB08,TSSASKKL12,P10,PF14}.  In this method,
galaxy  simulations are run to determine the gas response to a
bar pattern speed with a bar potential that has been determined from
near-infrared imaging.  The resulting simulations are compared with optical
observations.  The simulation that matches the observational morphology
closest determines the bar pattern speed.  This is considered an indirect
method for determining the pattern speed of the bar.

Another method for determining the CR location was introduced by \citet{C93}.
In this method, kinematic data along the minor axis of a spiral galaxy
shows a change in sign of the radial velocity across spiral arms.  This
change occurs at the location of the CR.  This has been investigated by
several authors \citep[e.g.][]{CA97,F01,F14}, but once again, the requirement
for spectroscopic data means that it is expensive in terms of telescope
time.

The {\em potential density shift method} can also be used to determine the
location of CR.  In this method, the pattern speed and the corotation radius
are determined using the azimuthal phase shift between the potential and the
density wave pattern \citep{ZB07}.  An issue with this method is that it
frequently uncovers ``superfast'' bars, where the corotation radius is
smaller than the length of the bar.  Under these circumstances,
the stability and morphology of the orbits in the corotation region do not
allow a bar to exist there \citep{C80}.

These methods have revealed that most bars have ``fast'' pattern speeds with
the ratio of the bar pattern speed to the bar length falling in the range
$1.0 \leq \mathcal{R} < 1.4$.  Some galaxies appear to harbor ``slow'' bars
(i.e., where $\mathcal{R} \geq 1.4$), but these seem to be the exception
rather than the rule.

In this paper, we concentrate on a method for determining the location of CR
that was first described by \citet[][hereafter PD]{PD97}.
\citet{SSTP15} used the PD method to show that the CR radius
of bars in spiral galaxies can be determined using imaging data alone.
\citet{SSTP15} used the PD method to determine
the corotation radii for 17 galaxies that already had CR measurements
determined by other methods in the literature and showed that the PD method
reproduced the same results.  They then went on to apply the method to another
50 barred spiral galaxies.

\begin{figure}
  \includegraphics[width=0.5\textwidth]{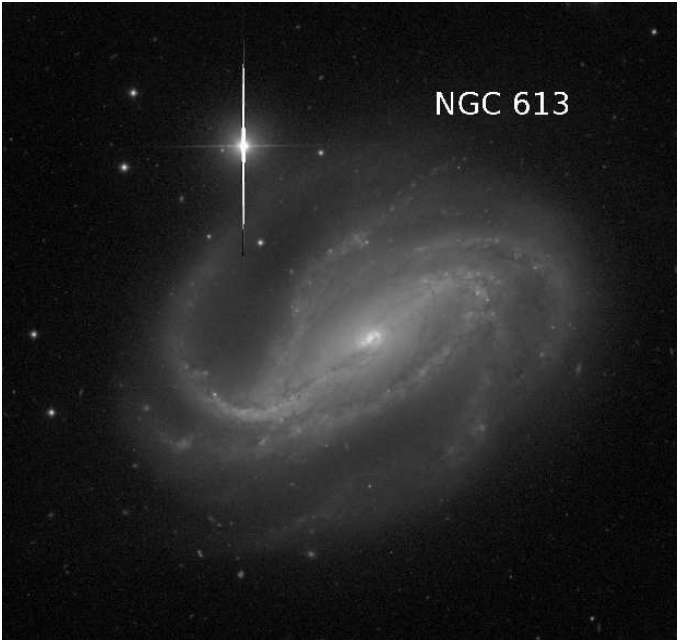}
  \caption{True color composite image of NGC 613.}
  \label{fig1}
\end{figure}

In this paper, we apply the PD method to NGC 613.  The PD method
in this galaxy finds two phase crossings.  The outermost phase crossing
probably corresponds to the corotation radius of the bar.
We also use our imaging data to highlight some nuclear structure to try and
understand the inner phase crossing revealed by the PD method.
We suggest that the inner phase crossing may be
indicative of nuclear activity in NGC 613
that has been revealed elsewhere in the literature.  For example
\citet{DGK17} and \citet{MSN18} both study star formation in the nuclear
region of NGC 613.  In their work, they find nuclear star formation
in a circum-nuclear disk extending out to the location of the nuclear phase
crossing that we have found here.

NGC 613 is classed as a SB(rs)bc galaxy \citep{RC3}.  It has a
heliocentric recession velocity of $V_{\rm rec}=1481$ km s$^{-1}$ and it has
a nucleus that is classified as having both Seyfert and H{\tt II} activity
\citep{VV06}.

\section{Data and Methods}

\subsection{Data Description}

For this project, we used $BVRI$ images of NGC 613
that were taken as part
of the Carnegie-Irvine Galaxy Survey \citep[][hereafter CGS]{Ho11}.  The
observations were taken at the 2.5-m du Pont Telescope at the Las Campanas
Observatory with the 2048$\times$2048 Direct CCD Camera on 4 April 2006
in photometric conditions.
The exposure times were 2$\times$360 s in $B$, 2$\times$180 s in $V$,
2$\times$120 s in $R$, and 2$\times$180 s in $I$.  The pixel scale for the
CCD camera is 0.259 arcsec/pixel, resulting in a field of view of
$8.85^{\prime} \times 8.85^{\prime}$.  The limiting surface brightness in
each waveband is 27.5, 26.9, 26.4, and 25.3 mag arcsec$^{-2}$ in the $B$, $V$,
$R$, and $I$ bands respectively.  A combined true-color image of
NGC 613 is shown in Figure \ref{fig1}.

\subsection{Mask creation}

In order to remove foreground objects (e.g., stars) and/or background galaxies
from the $BVRI$ images of NGC 613,
we used the Source Extractor routine
\citep{BA96}.  This routine was used to determine the location and extent
of foreground and background objects.  Source Extractor creates a FITS file
which was then manipulated in {\tt IRAF} to create a mask file.

\subsection{Image deprojection}

The Fourier transform code that we use only works on deprojected galaxy images.
In order to deproject galaxies, we assume that the outermost isophotes of
spiral galaxies are intrinsically circular in shape.  To do this, we use the
{\tt IRAF} task {\tt ellipse}.  This routine fits ellipses to isophotes in an
interative method described by \citet{J87}.  Using this routine, we derive the
ellipticity and position angle (PA) of the outermost isophotes of
NGC 613.
Deprojection of the galaxy images was then performed by rotating each image
through an angle equal to the PA of the major axis.  The images were then
stretched along the x-axis by an amount determined by the ellipticity of the
outer isophotes.  For each waveband, the same PA and ellipticity was used.
The final image was then inspected by eye to verify that the outermost
isophotes all appeared circular in shape.

\subsection{Determination of bar radius}
\label{rbar}

The deprojected image of the galaxy was viewed in the $I$ band and the
approximate visual end of the bar was recorded.  Then, the method described
in \citet{W95} was used to find the bar end.  Following this method, the PA
and ellipticity were plotted as a function of ${\sqrt{r}}$ as this facilitates
detection of details in the inner regions of images.  These plots can also
be used to highlight bar structure (see Figure 1 in \citet{W95}).  In a plot
of ellipticity as a function of $\sqrt{r}$, a bar shows up as a morphological
feature with a high ellipticity, and the end of the bar is highlighted by a
sudden change to the minimum ellipticity, $e_{\rm min}$.  Similarly, the PA
changes rapidly at the bar end.

\subsection{Determination of corotation radii}
\label{CR}

\begin{figure*}
  \centering
  \includegraphics[width=0.9\textwidth]{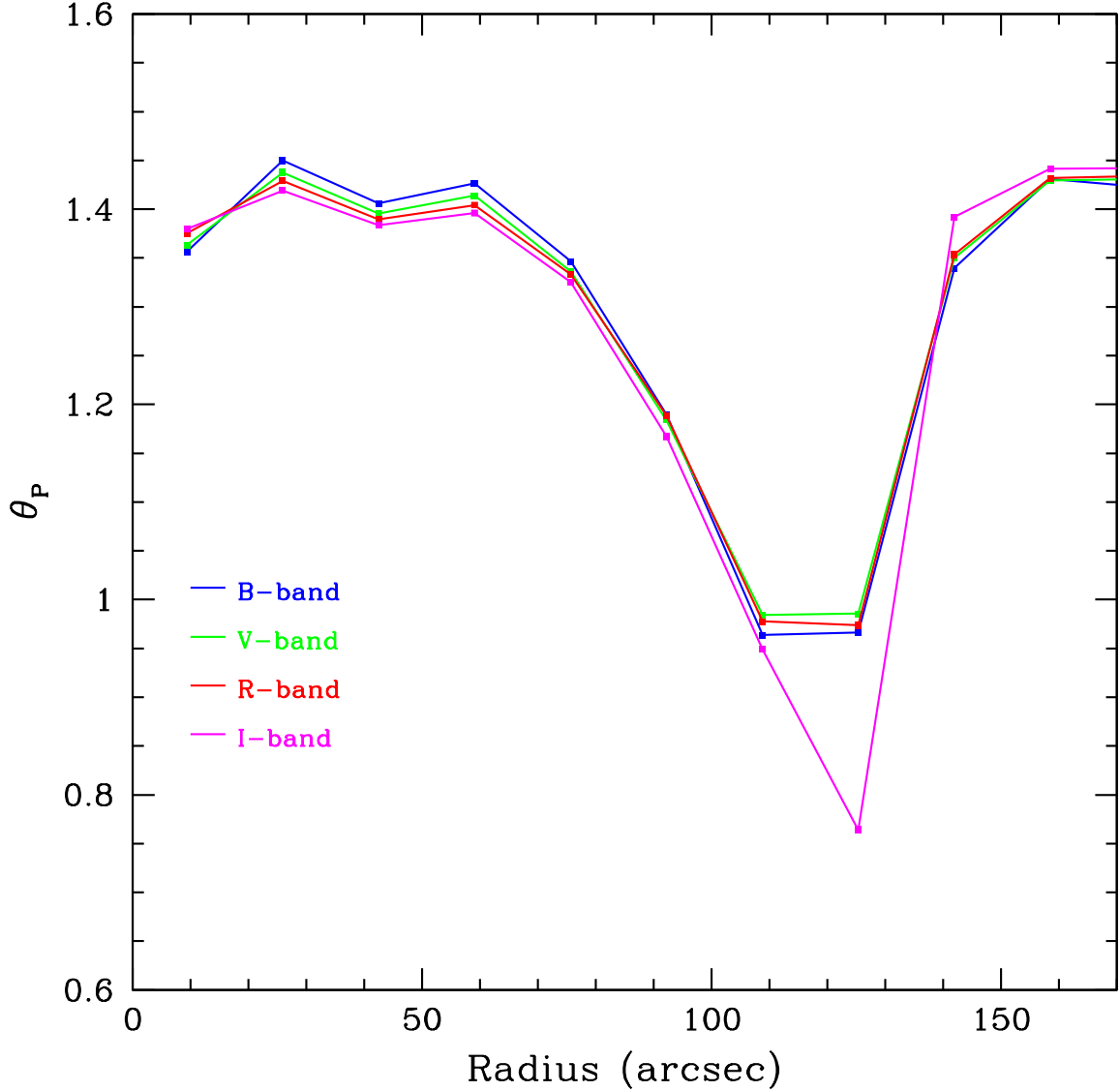}
  \caption{Phase angle as a function of radius for NGC 613.}
  \label{fig2}
\end{figure*}

\begin{figure*}
  \includegraphics[width=1.0\textwidth]{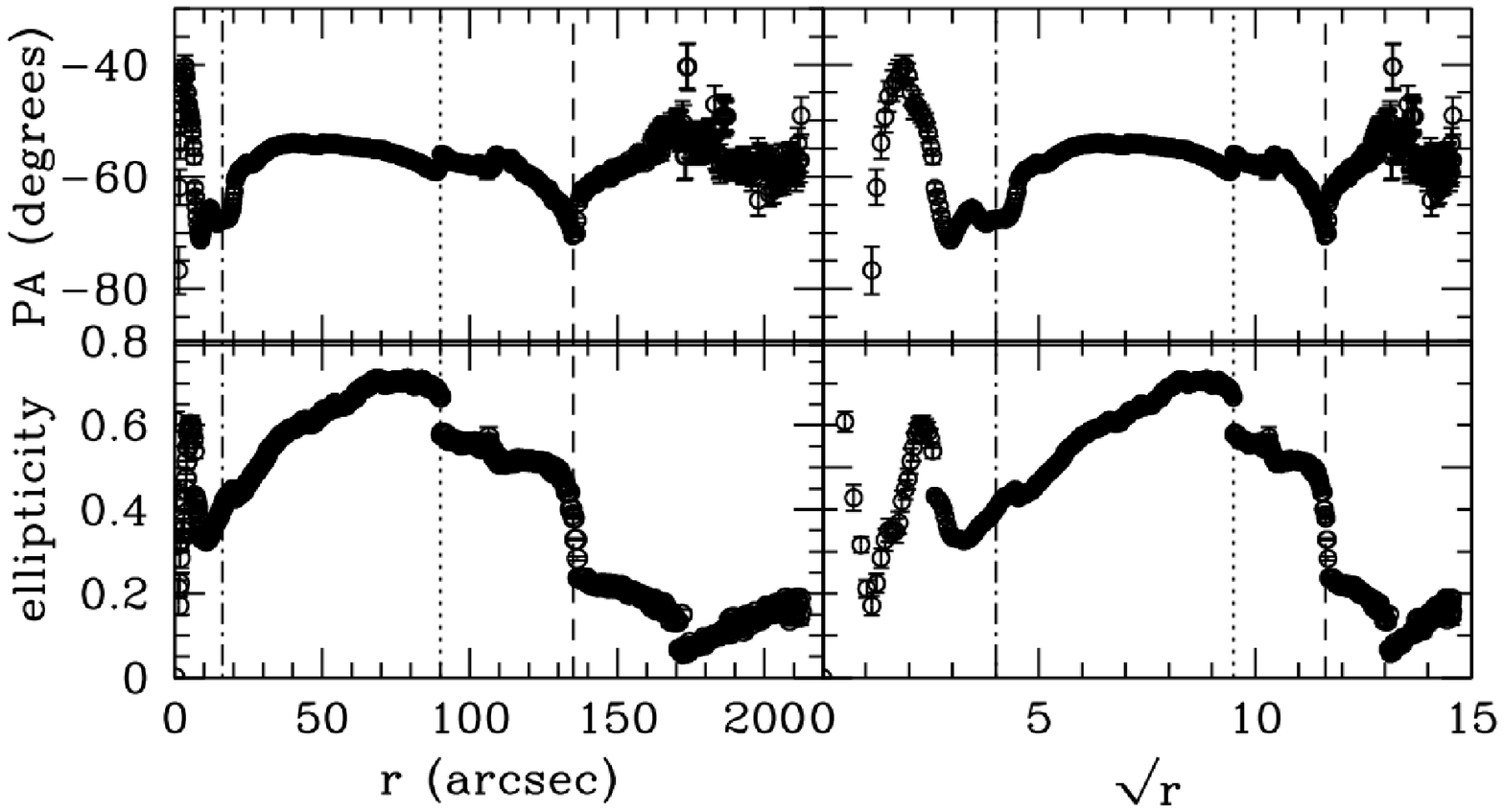}
  \caption{Position angle of the major axis and ellipticity as a function of radius for isophotes of NGC 613 as determined from ellipse fitting.  {\em Top left}: Position angle versus $r$; {\em Top right}: Position angle versus $\sqrt{r}$; {\em Bottom left}: Ellipticity versus $r$; {\em Bottom right}: Ellipticity versus $\sqrt{r}$.  In all panels the vertical lines represent the location of the inner corotation radius (dot-dashed line), the bar end (dotted line), and the outer corotation radius (dashed line).}
  \label{fig3}
\end{figure*}

The method we use in this paper for determining corotation radii is described
by \citet{SSTP15}, who first applied the method to a large sample of spiral
galaxies.  It is based upon the idea that color profiles of galaxies can be a
useful tool in studying spiral structure \citep{BC90}.  They showed that phase
shifts in the profiles of the $B$ and $I$ band spiral structure can highlight
the locations of star-forming regions.  PD then built upon this by suggesting
that the phase crossings of $B$ and $I$ band light indicates the location of a
corotation radius.  PD describes the photometric method for determing
corotation radii, and we refer the reader to this paper as well as
\citet{SSTP15} for a detailed discussion of how the method works and its
successes.

In summary, the PD method uses a code that creates azimuthal profiles of
galaxies and performs Fourier transforms on them.  The image is first divided
into 180 azimuthal sections each 2$^{\circ}$ wide, and 360 radial sections
each 1 pixel wide.  An azimuthal profile was created at each radial division.
The code then runs a Fourier transform for each of the azimuthal profiles,
i.e.,
\begin{equation}
  F_{2}(r)=\int_{-\pi}^{\pi}I_{r}(\theta)e^{-2i\theta}{\rm d}\theta
\end{equation}
where the associated phase angles are given by
\begin{equation}
  \theta(r)=\tan^{-1}\frac{{\rm Re}[F_{2}(r)]}{{\rm Im}[F_{2}(r)]}
\end{equation}
where ${\rm Re}[F_{2}(r)]$ and ${\rm Im}[F_{2}(r)]$ are the real and imaginary
parts of the complex Fourier coefficient, $F_{2}(r)$.  The phase angle as a
function of radius for each waveband was then plotted on the same graph (see
Figure \ref{fig2}).  Locations where the phase angles intersect are indicative
of corotation radii.

As with any method, the PD method does have its limitations.  One 
may think that dust attenuation may be a factor because we are using
visible light images.  However, the main reason this method works is because
at CR, the bluer star formation regions and the redder dust lane regions
flip across the spiral arm at the CR location.  As most of the dust attenuation
occurs within the dust lanes, the PD method actually relies, to some extent,
on this dust attenuation.  The main limitation with the
PD method actually occurs when multiple phase crossings occur, which can
lead to misinterpretations of such results.  Furthermore, in some cases
multiple phase crossings are seen within just a few arcseconds of each
other.  In these cases, authors often refer to this as a corotation region
\citep[see. e.g.][]{SSTP15}, although it is unclear why multiple crossing would
occur in such a small region of the disk of a galaxy.

\subsection{Calculation of the dimensionless quantity $\mathcal{R}$}

Here, we calculate
the following dimensionless parameter, i.e., the ratio of the CR radius to the
bar radius,
\begin{equation}
  \mathcal{R}=\frac{R_{\rm CR}}{R_{\rm bar}}
\end{equation}
where $R_{\rm CR}$ is the corotation radius determined in section \ref{CR} and
$R_{\rm bar}$ is the bar radius as determined in section \ref{rbar}.  If
$\mathcal{R}\geq 1.4$, the bar is defined as a ``slow'' rotator and if
$\mathcal{R}<1.4$, the bar is ``fast'' \citep{DS00}.  \citet{A14}
argues that $\mathcal{R}$ is
a good method for standardizing comparisons of bar pattern speeds in
different galaxies.

\subsection{Creation of structure and color maps}

To enhance the observable dust and star formation structure in our images
of NGC 613, we have chosen to create structure maps, a technique
originally developed by \citet{PM02}.  Mathematically, the structure map is
defined as
\begin{equation}
S=\left[\frac{I}{I \otimes P}\right] \otimes P^{t}
\end{equation}
where $S$ is the structure map, $I$ is the image, $P$ is the point-spread
function (PSF), $P^{t}$ is the transform of the PSF, and $\otimes$ is the
convolution operator.  Structure maps enhance structure variations on the
smallest resolvable scale of the image.  In the case of our optical imaging
data of NGC 613, the seeing in all four wavebands is $\sim0.5$
arcsec.

\section{Results and Discussion}

Figure \ref{fig2} shows the phase angle of the spiral structure as a function
of radius for all four wavebands, $B$, $V$, $R$, and $I$.  The locations on
this plot where all four wavebands intesect (a phase crossing) should denote
the 
location of corotation radii.  For NGC 613,
there are two phase crossings
highlighted in Figure \ref{fig2}, an inner phase crossing at
$r_{\rm CR1}=16\pm8$ arcsec and an outer corotation radius at
$r_{\rm CR2}=136\pm8$ arcsec.  The main source of the error in determining
these phase crossings is the frequency at which the phase angle is sampled.

For NGC 613, \citet{RSL08}
found a corotation resonance at $126.2\pm14.6$
arcsec using sticky particle simulations.  This is in fairly good agreement
with the location we determine for the outer corotation radius, i.e.,
$r_{\rm CR2}=136\pm8$ arcsec.

In order to determine the ratio of the outer corotation resonance to the
bar length, we now need to determine the bar length.  Figure \ref{fig3} shows
both position angle and ellipticity as a function of radius.  In these plots,
the end of the bar is shown as a sudden change from a maximum ellipticity.
Often (but not always) this is accompanied by a change in position angle.
From this analysis, we estimate the bar length as $R_{\rm bar}=90.0\pm4.0$
arcsec.  At this radius a sudden change can be seen in ellipticity, from a
maximum of $e_{\rm max}\simeq 0.7$, and a small but sudden change can also
be seen in the position angle at this radius.  This is slightly larger than
the bar length reported in \citet{RSL08} who found $R_{\rm bar}=78.6\pm3.2$
arcsec for their sticky particle simulations.  For their bar lengths, which are
used as inputs to the sticky particle simulations, \citet{RSL08} used a
modified version of the method used by \citet{E04,E05} to calculate the
size of the bar.

\begin{figure*}
  \includegraphics[width=0.45\textwidth]{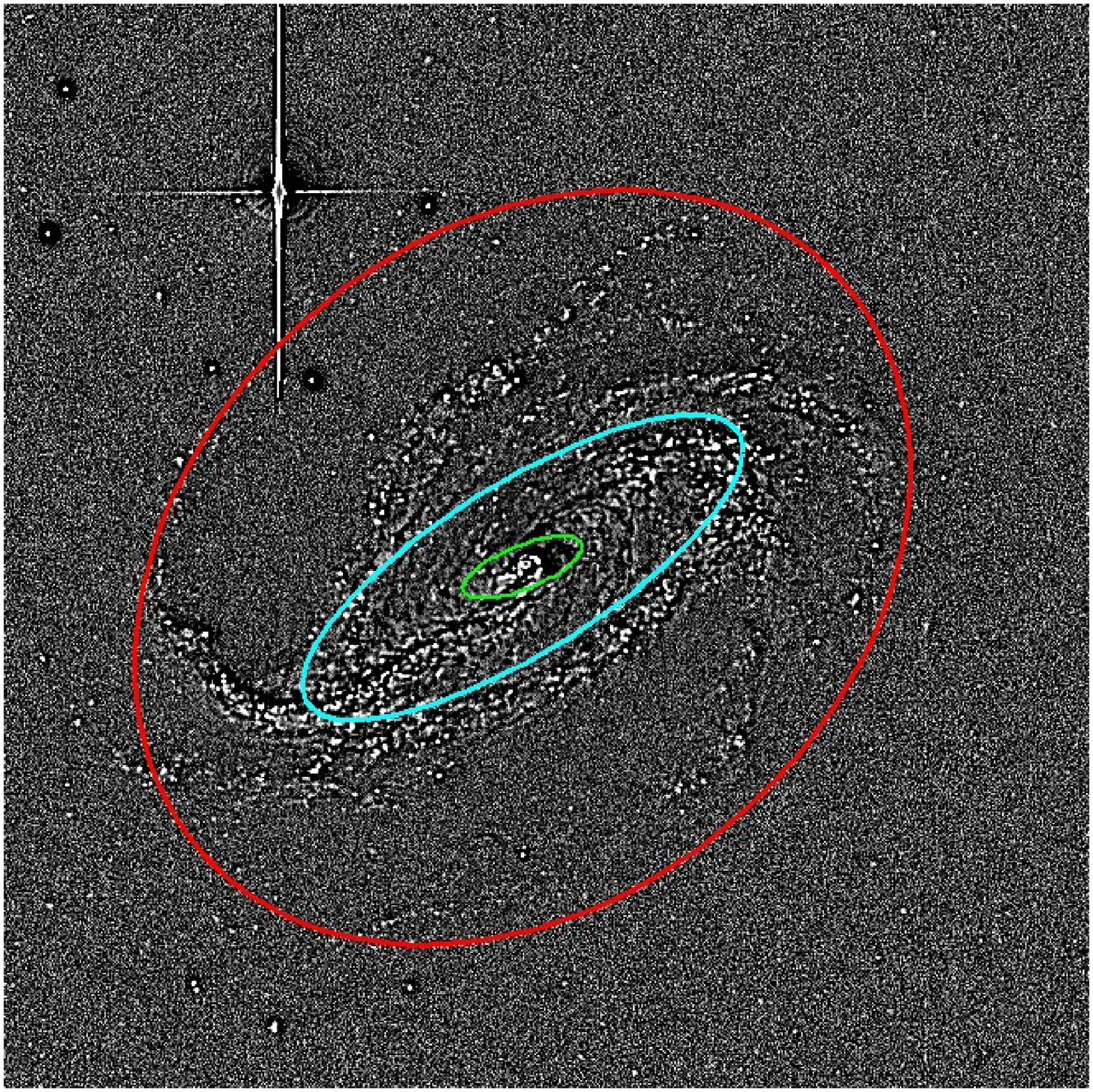}
  \includegraphics[width=0.45\textwidth]{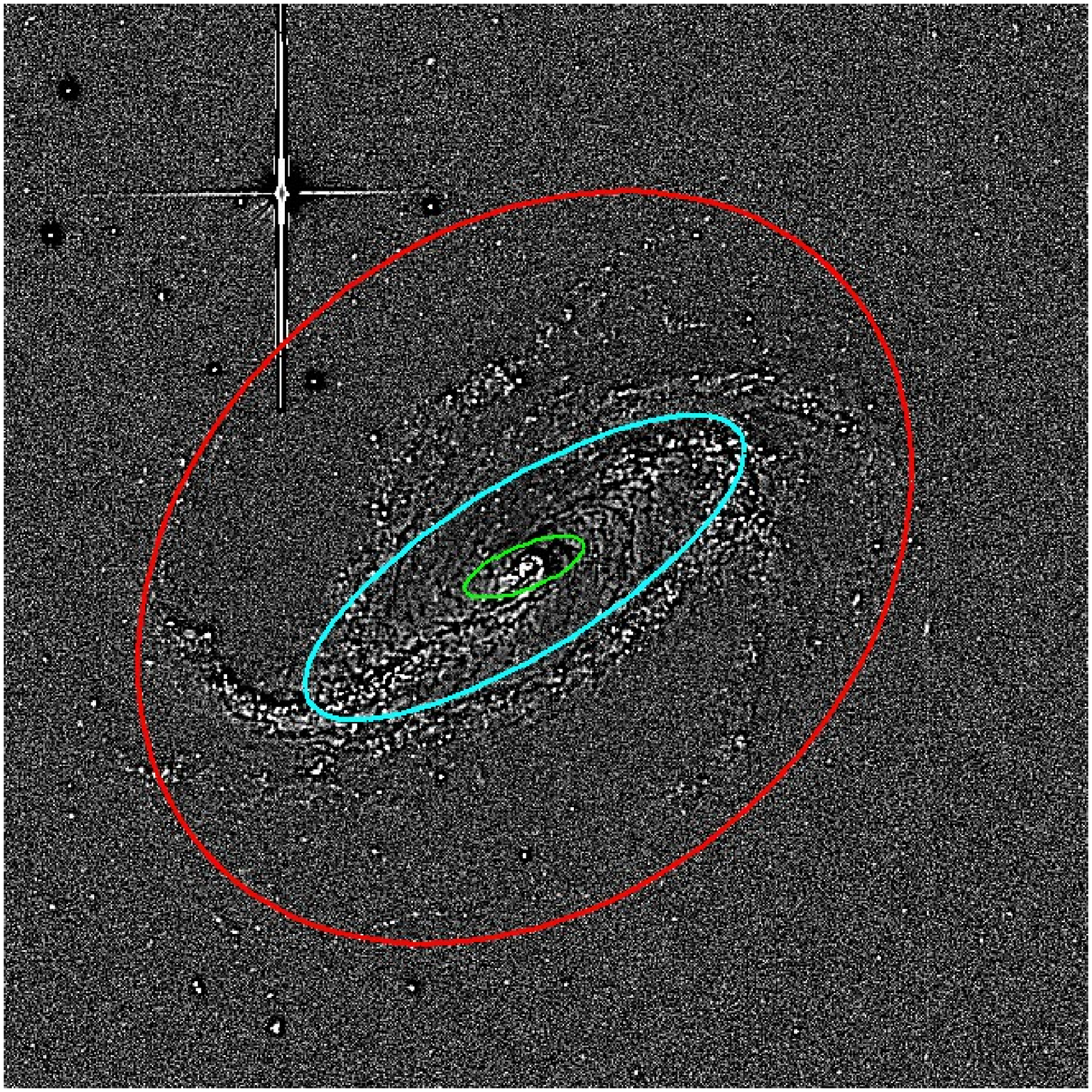}
  \includegraphics[width=0.45\textwidth]{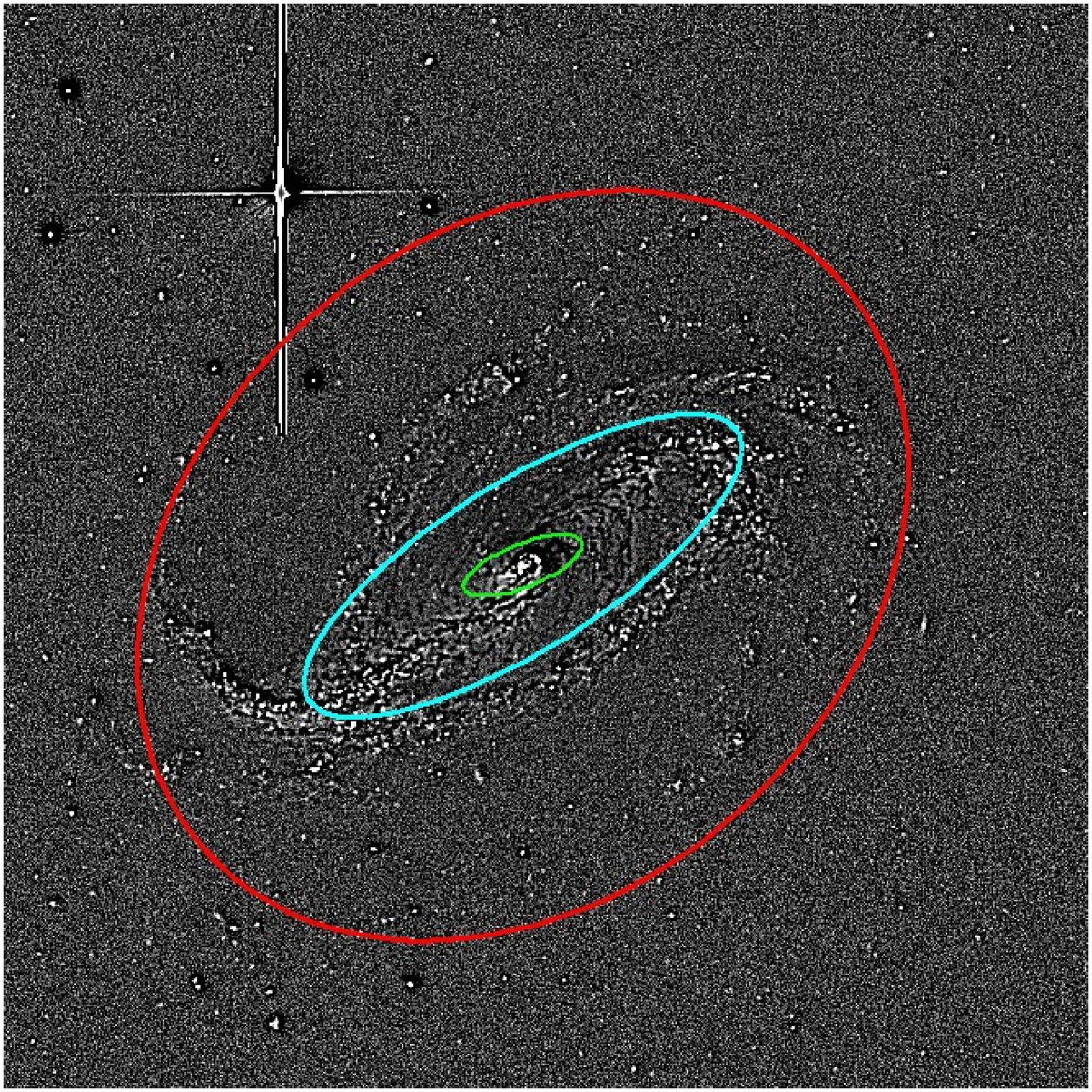}
  \includegraphics[width=0.45\textwidth]{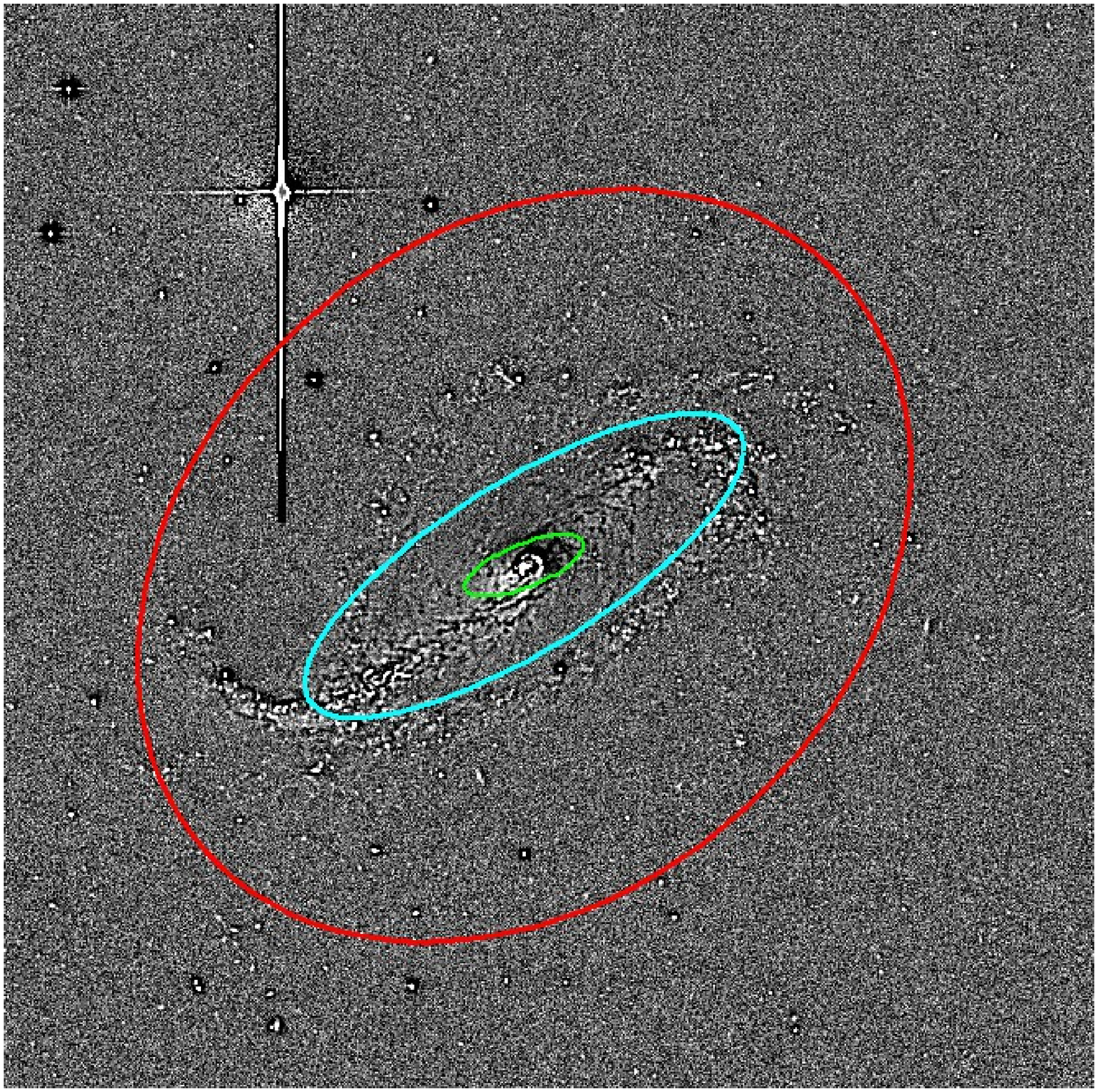}
  \caption{Structure maps of the $B$ band image of NGC 613 (top left), the $V$ band image (top right), the $R$ band image (bottom left) and the $I$ band image (bottom right).  In all images, the location of the bar end (cyan ellipse), the inner phase crossing (green ellipse), and the outer CR (red ellipse) are shown.  Each image is $360^{\prime\prime} \times 360^{\prime\prime}$ on the side.}
  \label{fig4}
\end{figure*}

\begin{figure*}
  \includegraphics[width=3.2in,height=1.81in]{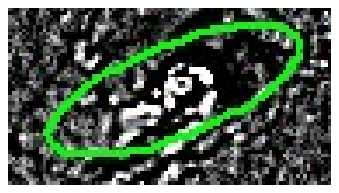}
  \includegraphics[width=3.2in,height=1.81in]{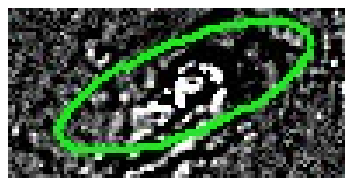}
  \includegraphics[width=3.2in,height=1.81in]{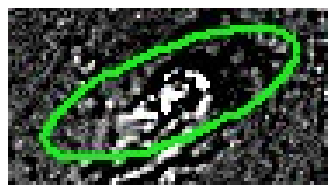}
  \includegraphics[width=3.2in,height=1.81in]{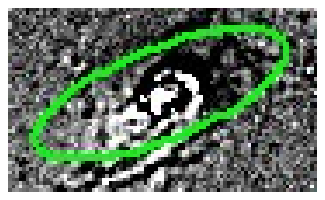}
  \caption{The same structure maps as in Figure \ref{fig4} but zoomed in on the central region with the inner phase crossing shown as a green ellipse.  Each image is $23^{\prime\prime} \times 13^{\prime\prime}$ on the side.}
  \label{fig4a}
\end{figure*}

Given the location of the outer corotation radius, $R_{\rm CR2}=136\pm8$ arcsec,
and the bar length, $R_{\rm bar}=90.0\pm4.0$ arcsec, that we have determined
here, we calculated the ratio of the outer corotation resonance to the
bar length
$\mathcal{R}=R_{\rm CR2}/R_{\rm bar}=1.5\pm0.1$.  For a comparison, \citet{RSL08}
derive a bar pattern speed of $\mathcal{R}=1.61\pm0.24$ from their sticky
particle simulations.  Our estimate is in agreement with their determination
of the bar pattern speed within the uncertainties.  Also, both our estimate
and that of \citet{RSL08} are consistent with $\mathcal{R}=1.4$, which is the
transition from slow ($\mathcal{R}\geq 1.4$) to fast ($\mathcal{R}<1.4$) bar
pattern speeds.  We therefore consider the bar in NGC 613 to have
a slow pattern speed.


As highlighted in Figure \ref{fig2}, we have found an inner phase
crossing at a radius of $R_{\rm CR1}=16\pm8$ arcsec.
One potential explanation for this inner phase crossing is that NGC 613
has an inner bar, or a bar within a bar \citep[e.g.][]{MS00,MA08},
in which case this phase crossing represents the corotation resonance of this
morphological feature.
However, in the position angle and ellipticity profiles in Figure 3, a bar
should show
up with approximately constant position angle and with an ellipticity that
reaches a high ellipticity and then suddenly changes to a minimum.  While
this is clearly the case for the large-scale bar in NGC 613, no such
indication exists in these profiles for a nuclear bar.  It is possible
that the data presented here, simply do not have the resolution to highlight
a nuclear bar, but such a feature has not been seen elsewhere in the literature
for NGC 613.  As a result, we have chosen not to focus on this explanation,
and instead focus on other forms on nuclear activity that is known to exist
in NGC 613.

\begin{figure*}
  \includegraphics[width=0.45\textwidth]{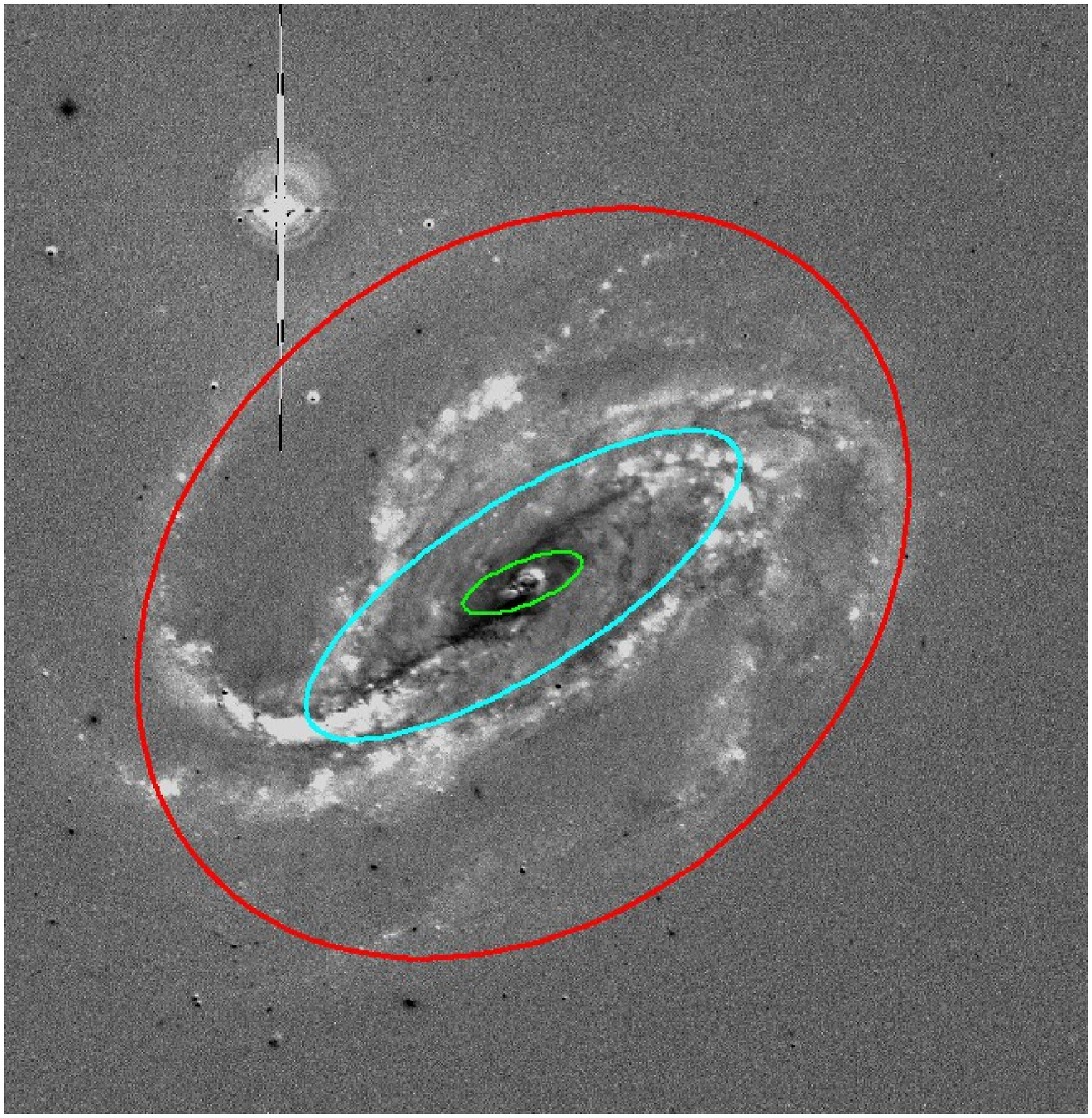}
  \includegraphics[width=0.45\textwidth]{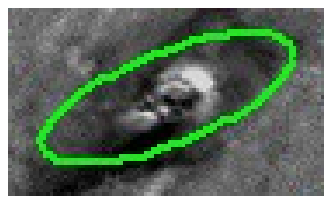}
  \caption{$B-R$ color map of NGC 613.  The overlaid ellipses indicate the location of the inner phase crossing (green), the outer CR (red), and the bar end (cyan). {\em Left Panel}: $360^{\prime\prime} \times 360^{\prime\prime}$ $B-R$ color image showing the whole galaxy with the inner phase crossing, bar region and outer CR.  {\em Right panel}:  $23^{\prime\prime} \times 13^{\prime\prime}$ zoomed $B-R$ color image of the galaxy out to the inner phase crossing.  In these color maps, the brighter areas are blue star-forming regions and the dark areas are red dusty regions.}
  \label{fig5}
\end{figure*}

The presence of an inner phase crossing may be consistent
with nuclear activity in NGC 613,
which is a low-luminosity AGN with prominent
radio jets \citep[e.g.,][]{GA09,HJ92}.  In order to study this in more
detail, we have created structure maps of NGC 613 (see Figures \ref{fig4}
and \ref{fig4a}) and
a $B-R$ color map (see Figure \ref{fig5}).

Both the structure maps (in $B$, $V$, $R$, and $I$) and the $B-R$ color map
show evidence of a partial ring of star formation at a radius of $\sim 4$
arcsec, which has been highlighted in previous works \citep[e.g.,][]{MSN18}.
Beyond this radius, the structure maps and $B-R$ color maps show no other
evidence of nuclear structure in NGC 613.
The large-scale bar and spiral
structure can clearly be seen in the outer parts of all of the structure
maps and the $B-R$ color map.

The partial circum-nuclear star formation ring shown in Figures \ref{fig4}
and \ref{fig5}, was also recently studied by \citet{MSN18}.  They studied
the central circum-nuclear disk and star-forming ring with the Atacama
Large Millimeter Array (ALMA) in [C{\tt I}], $^{13}$CO, and C$^{18}$O.  Their
observations place the star-formation ring at a radius of $\sim4$ arcsec (see
their Figure 1), well within the inner phase crossing that we have found.
Furthermore, \citet{DGK17} studied the AGN activity in NGC 613
using H$\alpha$,
H$\beta$, [N{\tt II}], [S{\tt II}], and O[{\tt III}] emission lines, and
they find evidence of star formation within the inner phase crossing,
at $R_{\rm CR1}=16\pm8$ arcsec, that we have found here (see their Figure 3).

The left panel of Figure \ref{fig5} also highlights that the major part
of the dominant two-armed spiral structure in NGC 613 is included inside
the outer corotation radius at $r_{\rm CR2}=138\pm8$ arcsec.  This is
consistent with the result of \citet{KSR03} who located corotation resonances
by comparing simulated gas density dsitributions for different spiral patterns
and compared them to observed disk morphology for a sample of five galaxies.
In all five galaxies, \citet{KSR03} found that the corotation resonance was
also located near the end of the dominant spiral pattern, at an average
galactocentric radius of $h=3.04\pm0.49$, where $h$ is the exponential disk
scalelength.

Nonlinear orbital modeling and some
N-body simulations also place corotation resonances in a similar location
\citep{P91,PK99}.  In the response models of \citet{PT17}, they find two
sets of spirals, which appear to be caused by different dynamical mechanisms.
The structure inside corotation appears to be supported by regular orbits
and this appears to be consistent with above theoretical studies.
Outside corotation, the spirals in the \citet{PT17} models appear to be
associated with ``chaotic spirals'' in both the stellar and the gaseous disk.
The two spirals meet near the unstable Lagrangian points of the system.
Our finding that the outer corotation resonance appears to be located near
the end of the dominant spiral pattern is, therefore, consistent with the
results of several dynamical simulations, and could be more evidence in favor
of the modal theory of spiral structure.

\section{Conclusions}

The main three conclusions of this paper are now summarized as follows:

\begin{itemize}

\item Using the PD (or multi-band photometric) method, we have determined that NGC 613 has an outer corotation radius at $R_{\rm CR2}=136\pm8$ arcsec.  Given our determination of the bar length of $R_{\rm bar}=90.0\pm4.0$, we find a relative bar pattern speed of $\mathcal{R}=1.5\pm0.1$, which is consistent with that found using other methods \citep{RSL08}.  This suggests that NGC 613 has a bar with a slow pattern speed.

\item We have also found that NGC 613 has an inner phase crossing, although this is likely not related to a corotation resonance.  Instead, we suggest that this inner phase crossing is indicative of nuclear activity, in particular nuclear star formation.  Previous works have indicated the presence of a star-forming ring and a star-forming nuclear disk in NGC 613 \citep{PM06,FRB14,DGK17,MSN18}.  These works highlight a nuclear star-forming disk that extended out to $\sim12-15$ arcsec.  This is consistent with the location of our inner phase crossing radius at $R_{\rm CR1}=16\pm8$ arcsec.  We therefore suggest that phase crossings may also be useful for detecting the limit of nuclear activity, particularly nuclear star formation.

\item We find that the outer corotation resonance, located at $R_{\rm CR2}=138\pm8$, is close to the end of the dominant two-armed spiral pattern.  On the surface, this appears to be consistent with the expectations from theoretical models and simulations \citep[e.g.][]{KSR03,B89a,B89b,B93,PT17}.  According to the theory of global modes \citep{B89a,B89b,B93,BL96}, the corotation radius should be located in the outer parts of the disk.  This is consistent with our findings.
  
\end{itemize}
  
It should be noted that nuclear rings of star formation are likely due to $x_{2}$ orbits in the bar region and they they migrate inwards over time \citep{RT03}.  Indeed, simulations of galaxies have shown that these nuclear rings of star formation have a dusty rings to their interiors \citep{RT03}, and this supports the idea that the ring is migrating inwards.  However, at the resolution of the images presented in this paper, it is unlikely that we can detect the interior dusty ring in NGC 613.  Instead, we believe that we have detected the extended disk reported by several authors \citep[e.g.,][]{MSN18}.

Finally, taking the results of \citet{TSSASKKL12}, \citet{SSTP15} and \citet{KSHP18} as well as the result for NGC 613 here, it seems that there is a phase crossing or corotation resonance in the outer part of the visible spiral structure whether or not there is a central bar.  \citet{SLBK10} have statistically shown that there is a correlation between bar strength and spiral arm amplitude out to a radius of $\sim 1.6 R_{\rm bar}$.  This suggests that either the spiral arms are a continuation of the bar mode or they are driven by the bar out to that radius.  Further out, spirals may be independent.  For NGC 613, we determine that the corotation radius is $R_{\rm CR2}=136\pm8$ arcsec $=(1.5\pm0.1)R_{\rm bar}$, which is inside the limit of $1.6R_{\rm bar}$ suggested by \citet{SLBK10}.  At least superfically, our result appears to be consistent with the result of \citet{SLBK10}.

\section*{Acknowledgements}

MSS wishes to thank the University of Minnesota Duluth for their support and
the Fund for Astrophysical Research.
The data presented in this paper were collected as part of the Carnegie-Irvine
Galaxy Survey (CGS), using facilities at Las Campanas Observatory, Carnegie
Institution for Science.  The optical data were reduced independently from
those presented in \citet{Ho11}.  The authors wish to thank the anonymous
referee for comments that significantly improved the content of this paper.











\bsp	
\label{lastpage}
\end{document}